\documentclass[aps,prl,twocolumn,showpacs,superscriptaddress, longbibliography]{revtex4-1}

\usepackage{xcolor, graphicx}
\usepackage{amssymb,amsmath, amsfonts, bm}
\usepackage{newtxtext,newtxmath}

\usepackage{dcolumn}
\usepackage[none]{hyphenat} 
\usepackage{tabularx}

\usepackage{verbatim}  

\begin{document}

\title{Antiferromagnetic quantum spin Hall states in iron halogenide}

\author{Qian Sui}
\email{These three authors contribute equally to this work.}
\affiliation{School of Physical Science and Technology, ShanghaiTech University, Shanghai 200031, China}
\author{Jiaxin Zhang}
\email{These three authors contribute equally to this work.}
\affiliation{School of Physical Science and Technology, ShanghaiTech University, Shanghai 200031, China}
\affiliation{Institute for Advanced Study, Tsinghua University, Beijing 100084, China}
\author{Suhua Jin}
\email{These three authors contribute equally to this work.}
\author{Yunyouyou Xia}
\affiliation{School of Physical Science and Technology, ShanghaiTech University, Shanghai 200031, China}
\author{Gang Li}
\email{ligang@shanghaitech.edu.cn}
\affiliation{School of Physical Science and Technology, ShanghaiTech University, Shanghai 200031, China}
\affiliation{\mbox{ShanghaiTech Laboratory for Topological Physics, ShanghaiTech University, Shanghai 200031, China}}

\begin{abstract}
It is widely known that quantum spin Hall (QSH) insulator can be viewed as two copies of quantum anomalous Hall (QAH) insulator with opposite local magnetic moments.
However, nearly every QSH insulator discovered so far is a nonmagnetic semiconductor.  
Due to the vanishing local magnetic moment of each copy, the QAH states only conceptually exist in these QSH insulators. 
In this work, we show a realistic construction of QSH states with finite local magnetic moment by staking bilayer QAH insulators. 
Our explicit construction benefits from an effective QAH model with a large topological gap and is further supported by a class of two-dimensional ferromagnetic materials. 
Our work not only validates the conceptual relationship of QSH and QAH but also provides an ideal material platform for realizing antiferromagnetic QSH state which is highly tunable between QAH and QSH states as a function of the number of layers. 
\end{abstract}

\maketitle

{\it Introduction.} -- 
The discovery of the mystery integer steps in the conductance of quantum Hall insulators~\cite{PhysRevLett.45.494} opens the gate to the previously unknown topological realm~\cite{RevModPhys.82.3045, RevModPhys.83.1057}, where each state differs from another by a topological property instead of symmetry.  
For the quantum Hall insulator, the topological character is known as Chern or TKNN number~\cite{PhysRevLett.49.405}, which deeply relates to the Berry phase of all occupied bands~\cite{Berry1984}.  
Haldane's model~\cite{PhysRevLett.61.2015} was the first model to realize quantum Hall states without Landau levels, which is now known as quantum anomalous Hall (QAH) insulators.  
The directional complex hopping introduces net magnetic flux in each hexagon of the honeycomb lattice. 
Different topological phases are monitored by the berry curvatures at the two independent Brillouin Zone (BZ) corner $\vec{K}$ and $\vec{K}^{\prime}$. 
The transition between states with different topology is controlled by the different signs of the curvature instead of symmetry. 
The quantum spin Hall (QSH) states were soon proposed as a superposition of two QAH insulators with equal but opposite magnetic moments~\cite{Kane2005}. 
While, despite the earlier proposal of QAH, QSH became the first topological state discovered in experiment without external magnetic field. 
Experimentally, HgTe/CdTe quantum well~\cite{hgte1, hgte2} was the first discovered QSH insulator which is controlled by the thickness of HgTe and, microscopically, by the inversion of the $\Gamma_{6}$ and $\Gamma_{8}$ states.  
In addition to the band-inversion mechanism, a graphene-type but of large gap QSH insulator~\cite{Reis287, PhysRevB.98.165146, PhysRevB.98.161407} with local spin-orbital coupling (SOC) has also been discovered recently, which even features spintronic application at room temperature. 

Although the QAH states are declared to exist in any QSH insulator, it is never shown explicitly to disassemble the so-far discovered QSH insulators to achieve QAH states. 
Instead, one has to dope topological insulators with dilute magnetic impurities to establish long-range ferromagnetic order through either surface RKKY-type interaction~\cite{PhysRevLett.102.156603} or van-Vleck mechanism~\cite{Yu61}. 
It has to be noted that breaking time-reversal symmetry (${\cal T}$) in such a way does not decouple the conceptual two copies of QAH states, but rather to invert one copy back to normal band order while still keeping the other one inverted.   
However, an ideal QSH insulator constructed by stacking two copies of QAH insulators would allow reverse engineering to get QAH states as well, which, ultimately, serves as the sufficient condition to validate the relationship of QSH and QAH states. 

In this letter, we explicitly study this problem by first extending the Haldane model to a two-orbital system with finite local SOC.
It demonstrates two distinct topological phases with $c=-1$ and $c=2$. 
We further show that this model nicely explains a class of two-dimensional ferromagnetic materials which are large-gap QAH material candidates holding strong potential for room-temperature application. 
After further extending this large-gap QAH model to a QSH model while keeping the local magnetic moments, we obtain a promising reversible routine to construct antiferromagnetic QSH insulators from QAH insulators that unbiasedly validates the relationship between them. 
The proposed stacking routine is further confirmed by a number of material candidates, which can switch between QAH and QSH states as a function of the number of layers.

\begin{figure}[htbp]
\centering
\includegraphics[width=\linewidth]{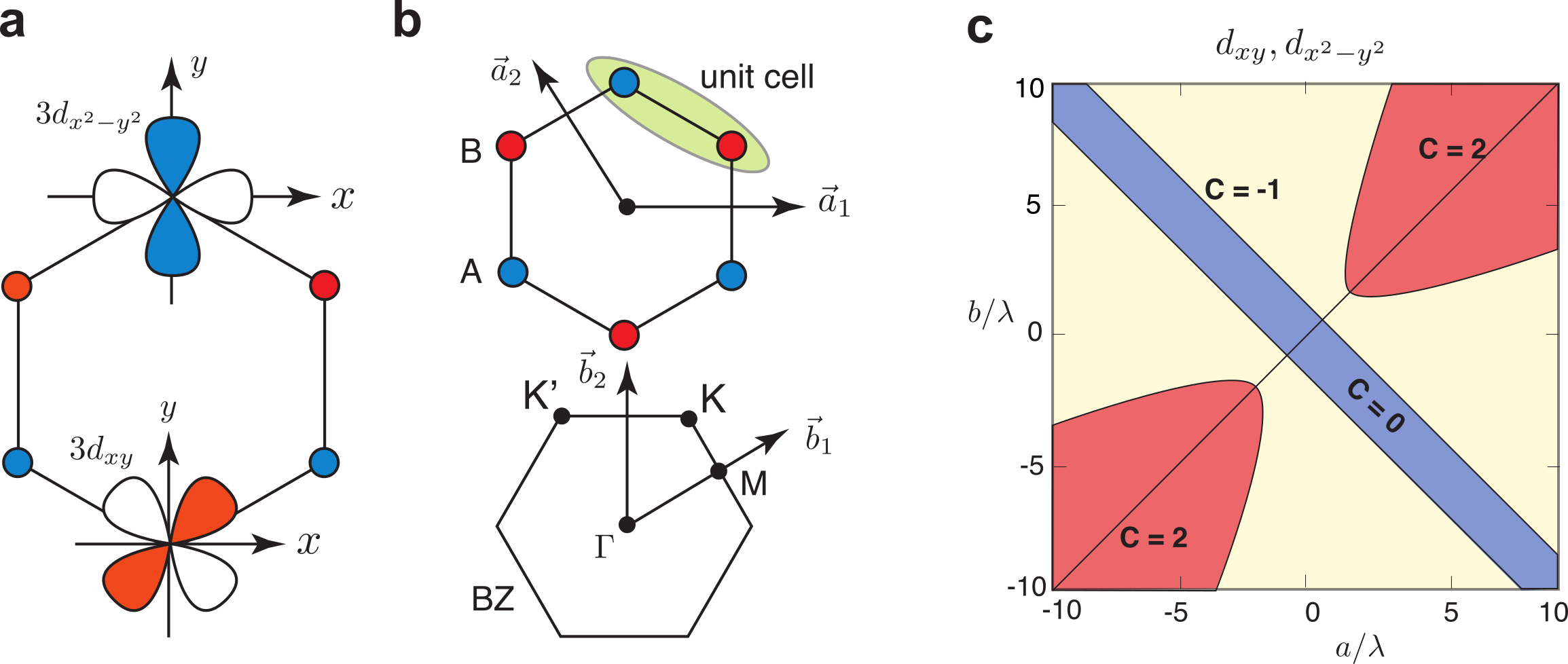}
\caption{Illustration of a two-orbital model on Honeycomb lattice and the corresponding phase diagram.  (a) $d_{xy}$-$d_{x^{2}-y^{2}}$ model. (b) The Honeycomb lattice and the corresponding BZ. The lattice and reciprocal vectors are defined as $\vec{a}_{1}=(1, 0)$, $\vec{a}_{2}=(-\frac{1}{2}, \frac{\sqrt{3}}{2})$, $\vec{b}_{1}=(2\pi, \frac{2\sqrt{3}\pi}{3})$, $\vec{b}_{2}=(0, \frac{4\sqrt{3}\pi}{3})$. (c) The phase diagram of the proposed model as a function of parameters $a/\lambda$ and $b/\lambda$.  }
\label{Fig1}
\end{figure}

{\it Results.} -- 
Here we first generalize Haldane's model to two orbital case.
Instead of the complex next-nearest neighbor hopping, we aim to get a complex coupling locally between the two-orbitals.
In this way, a bigger SOC and, consequently, a substantial increase of topological gap  can be expected. 

For this reason, we consider orbital combination $d_{xy}/d_{x^{2}-y^{2}}$. 
Here we use $t_{\alpha\beta}^{AB}$ to represent the hopping amplitude between orbitals $\alpha/\beta$ at the two inequivalent sites $A/B$. 
The general two-orbital tight-binding Hamiltonian on honeycomb lattice takes the following form:
\begin{equation}\label{Eq_tb}
H_{0} = 
\begin{pmatrix}
\epsilon_{\alpha}^{A} & 0 & \vline & h_{\alpha\alpha}^{AB} & h_{\alpha\beta}^{AB} \cr
0 & \epsilon_{\beta}^{A} & \vline & h_{\beta\alpha}^{AB} & h_{\beta\beta}^{AB} \cr
\hline
h_{\alpha\alpha}^{BA} & h_{\alpha\beta}^{BA}  & \vline & \epsilon_{\alpha}^{B} & 0 \cr
h_{\beta\alpha}^{BA} & h_{\beta\beta}^{BA} & \vline & 0 & \epsilon_{\beta}^{B}
\end{pmatrix}\;,
\end{equation}
where the diagonal matrix element $\epsilon_{\alpha/\beta}^{A/B}=\epsilon$ represents the orbital potential of $\alpha/\beta$ at site $A/B$.
They are taken as equal in our calculations, which represent the energy levels of the Dirac point.
They control the position of the Dirac linear crossing at $\vec{K}$ and $\vec{K}^{\prime}$ when there is no SOC. 
Off-diagonal $h_{\alpha\beta}^{AB}$ denotes the Hamiltonian element formed by hopping $t_{\alpha\beta}^{AB}$.
One simple way to correctly account for the orbital symmetry is to follow the Slater-Koster table~\cite{PhysRev.94.1498}. from which one can easily get $h_{\alpha\alpha}^{AB}=M_{1}(1+e^{ik_{1}})+M_{2}e^{-ik_{2}}$, $h_{\beta\beta}^{AB}=M_{3}(1+e^{ik_{1}})+M_{4}e^{-ik_{2}}$, $h_{\alpha\beta}^{AB}=\sqrt{3}M_{5}(1-e^{ik_{1}})$. 
Here, $k_{1}$ and $k_{2}$ are the fractions of reciprocal lattice vector $\vec{b}_{1}$ and $\vec{b}_{2}$.
The Hamiltonian is Hermitian as $h_{\alpha\beta}^{BA}=(h_{\alpha\beta}^{AB})^{*}$ and $h_{\beta\alpha}^{AB}=h_{\alpha\beta}^{AB}$. 
The corresponding coefficients $(M_{1}\cdots M_{5})$ are not independent and they are parametrized with the standard Slater-Koster integrals as $M_{1}=(3a+b)/4$, $M_{2}=b$, $M_{3}=(a+3b)/4$, $M_{4}=a$ and $M_{5}=(a-b)/4$ with $a=(3V_{dd\sigma}^{1}+V_{dd\delta}^{1})/4$ and  $b=V_{dd\pi}^{1}$. 

The proposed orbital combination is of special importance to the topological nature of the system, as it allows non-vanishing onsite SOC.  
Given the atomic SOC $\lambda\vec{L}\cdot\vec{S}$, it has a constant expectation value between the two orbital states proportional to strength $\lambda$. 
\begin{equation}\label{Eq_soc}
\langle d_{xy}|\lambda\vec{L}\cdot\vec{S} | d_{x^{2}-y^{2}}\rangle = 2i\lambda\;.
\end{equation}
This resembles the $p_{x}/p_{y}$ model of bismuthene~\cite{Reis287, PhysRevB.98.165146, PhysRevB.98.161407} where the large topological gap derived from the local SOC makes it a good material candidate for room-temperature spintronic application.  
By breaking inversion symmetry, one can further include Rashba-type SOC to the model, as done in the bismuthene $p_{x}/p_{y}$ model. 
In this work, we concentrate on the Hamiltonian $H_{qah}(k)=H_{0}(k)+\lambda\vec{L}\cdot \vec{S}$ and neglect other types of SOC for simplicity, which does not affect the topological nature of the models. 

\begin{figure}[htbp]
\centering
\includegraphics[width=\linewidth]{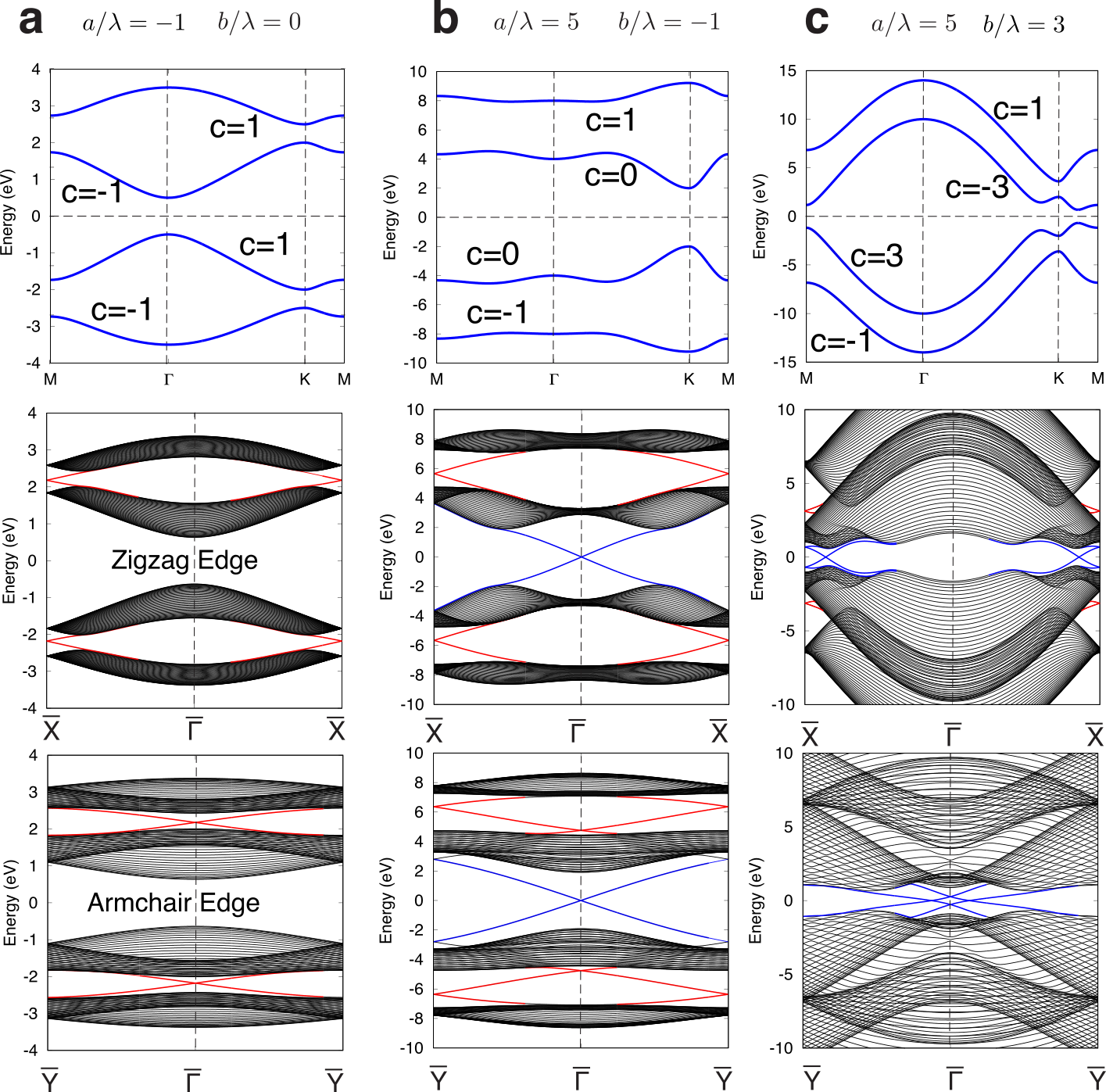}
\caption{The bulk and edge electronic structures of the $d_{xy}/d_{x^{2}-y^{2}}$ model in three topologically different phases: (a) $c=0$; (b) $c=-1$; (c) $c=2$. }
\label{Fig2}
\end{figure}

Eqns.~(\ref{Eq_tb}) and (\ref{Eq_soc}) consist our QAH Hamiltonian, whose generic phase diagram as a function of $a$ and $b$ is shown in Fig.~\ref{Fig1} (c). 
Chern number, the topological invariant charactering this spinless (or fully polarized) tight-binding model,  indicates a rich topological phase diagram. 
Except for the small regime around $a = -b$ and $a=b$, large areas of the parameter spaces host topologically nontrivial states and two distinct phases are discovered with $c = -1$ and $c=2$.
In Fig.~\ref{Fig2} we further display examples of all three topologically distinct phases with the representative band and edge electronic structures. 
We explicitly determine the Chern number for each band.
The phase diagram Fig.~\ref{Fig1}(c) was obtained from the sum of the Chern numbers of the two occupied bands.
From the bulk-boundary correspondence, Chern number indicates precisely the number of edge modes. 
Interestingly, even for the $c=0$ phase in Fig.~\ref{Fig1}(c), the topology of each band is nontrivial.
As shown in Fig.~\ref{Fig2}(a), despite the zero net Chern number of the two occupied bands, each band carries a non-zero topological invariant allowing the presence of the nontrivial edge mode.    
Thus, the gap between the first and the second bands, as well as the one between the third and fourth bands host topological edge states. 
From $c=0$ to $c=-1$ phases, the gap between the second and the third band closes and reopens at $\Gamma$-point. 
From $c=-1$ to $c=2$ phases, similar gap closing and reopening occurs at $M$-point. 
Both $\Gamma$ and $M$ are time-reversal invariant momenta of the honeycomb lattice, consequently, the topology of the model changes when the gap closes and reopens at these points. 
Note that, for the $c=2$ phase, along each edge there are two edge modes and they connect the valence with conduction bands around  $M$-point instead of $\Gamma$ along the zigzag edge. 

\begin{figure}[htbp]
\centering
\includegraphics[width=\linewidth]{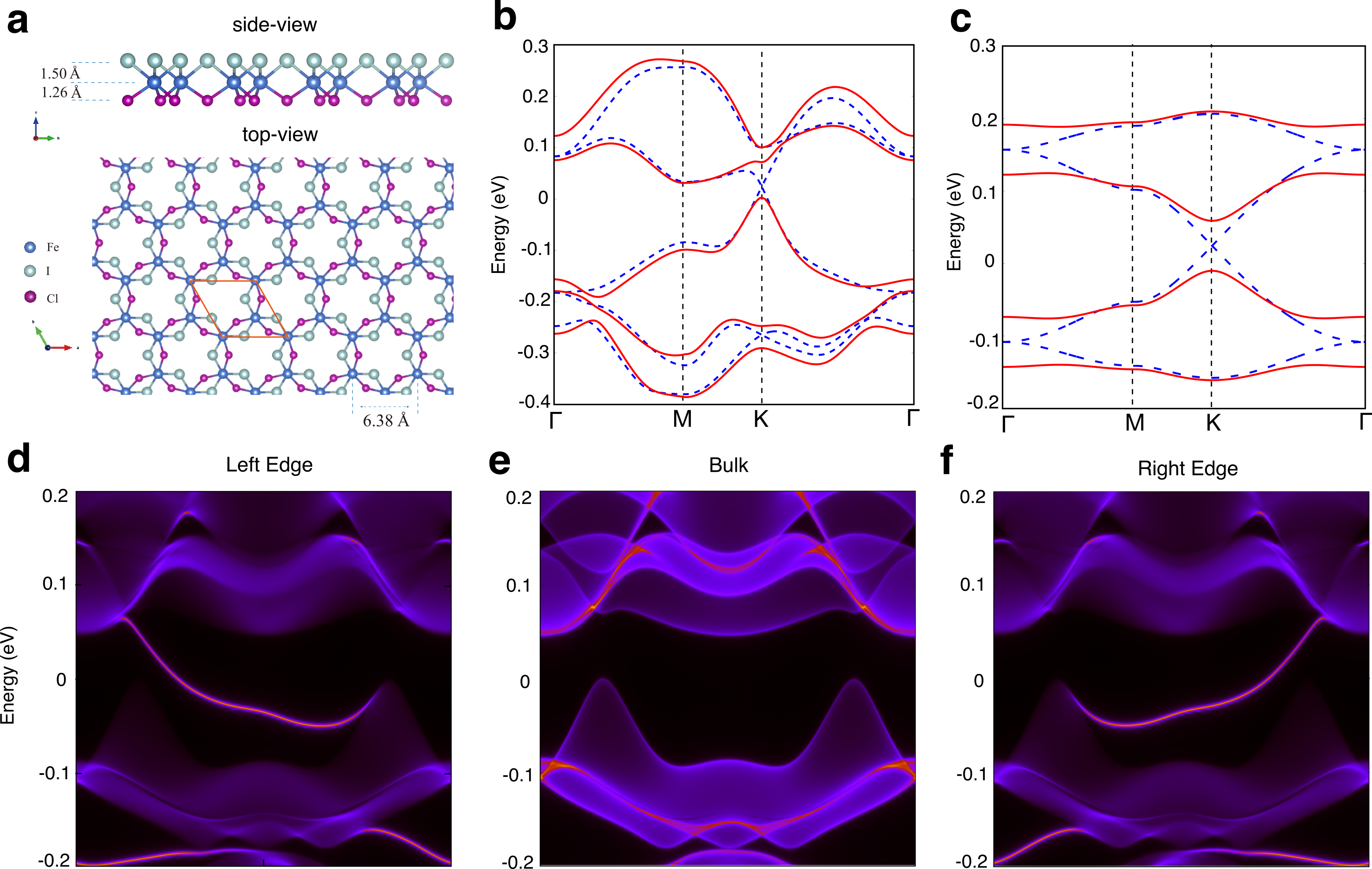}
\caption{(a) The structure model of monolayer Fe$_{2}$Cl$_{3}$I$_{3}$ in side and top views. The iron atoms form an effective honeycomb lattice. (b) The electronic structure with (red solid line) and without (blue dashed line) SOC. (c) The model bands with fitting parameters from Tab.~\ref{Tab1}. (e - f) The edge states calculations for the left, the bulk and the right edges, respectively. Both left and right edges are terminated with zigzag geometry.}
\label{Fig3}
\end{figure}

The proposed correlated QAH model nicely explains a class of two-dimensional ferromagnetic semiconductors represented by FeBr$_{3}$, which was recently proposed as a large-gap QAH insulator~\cite{zhang2017intrinsic}. 
While, we want to note that this is not the only material candidate which realizes the topological physics of the proposed tight-binding model.
On a honeycomb lattice, as long as the essential physics around the Fermi level is determined by $d_{x^{2}-y^{2}}/d_{xy}$ orbitals, the proposed model applies. 
With such a guidance, we manually replaced Br with other elements in group-VII without affecting the low-energy $d$-orbitals of iron.
Consequently, we found a class of large-gap QAH candidate materials.  
These include FeCl$_{3}$, FeI$_{3}$, Fe$_{2}$Cl$_{3}$Br$_{3}$,  Fe$_{2}$Cl$_{3}$I$_{3}$ and  Fe$_{2}$Br$_{3}$I$_{3}$, which all nicely resemble the $c=-1$ QAH phase of our model. 
By fitting our model to the density functional theory calculations and further requiring them to coincide at $\Gamma$ and $K$ points, we obtained the model parameters listed in Tab.~\ref{Tab1}. 
The corresponding electronic structure and the edge states can be found in Supplementary Information.  

In Fig.~\ref{Fig3} we show a representative material candidate Fe$_{2}$Cl$_{3}$I$_{3}$. 
As indicated by Fig.~\ref{Fig3}(a), chlorine and iodine layers stay below and above the iron layer. 
The local 3-fold rotation symmetry and the mirror symmetry are maintained, which is essential for reserving the linear Dirac band crossing at $\vec{K}$ and $\vec{K}^{\prime}$ on honeycomb lattice without SOC. 
Subsequently, the SOC opens a gap of size 69 meV as shown in Fig.~\ref{Fig3}(b), which is sufficiently large for room-temperature experiment and applications. 
The corresponding model electronic structure with/without SOC is shown in Fig.~\ref{Fig3}(c), which nicely captures the essential topology of the two occupied and two unoccupied states. 
The calculated Chern number is -1 for the two occupied bands. Individually, the four bands carry $c=-1$, $c=0$, $c=0$ and $c=1$, respectively. 
We further terminated the system with zigzag edge and determined the correspondingly edge states as shown in Fig.~\ref{Fig3}(e-f). 
As expected, compared to the bulk bands, both edges develop topologically nontrivial modes confirming this system as a QAH insulator. 
It should be further noted that, in addition to the topological edge states around the Fermi level, there also exists another QAH edge mode between the first and the second valence bands. 
Without SOC, the bulk electronic structure also displays a Dirac band crossing at $\vec{K}$ and $\vec{K}^{\prime}$ between these two bands. 
This exactly resembles the topological physics of the gap around the Fermi level. 
And this also agrees nicely with the model Hamiltonian in $c = -1$ phase, see Fig.~\ref{Fig2}(b) for the bulk and edge electronic structures.

\begin{table}
\centering
\setlength{\tabcolsep}{3.5pt}
\begin{tabular}{|c|c|c|c|c|c|c|}
\hline
   & FeCl$_{3}$ & FeBr$_{3}$ & FeI$_{3}$ & Fe$_{2}$Br$_{3}$I$_{3}$ & Fe$_{2}$Cl$_{3}$I$_{3}$ & Fe$_{2}$Cl$_{3}$Br$_{3}$ \cr
 \hline
 $a$  &  94.1 & 85.2 & 82 & 88 & 105 & 96.6  \cr
 \hline
 $b$ & -28.9 & -14.2 & -5.8& -7.9& -16.5&  -36.8  \cr
 \hline
 $\lambda$ &  17.5& 17.2 & 10.3& 14 & 17.3&  17.3  \cr 
 \hline
 $\Delta$ &  70 & 68& 41& 56& 69 & 69  \cr
 \hline
 $\epsilon$ & 16 & 20 & 27 & 34 & 24 & 17 \cr
\hline 
\end{tabular}
\caption{Model parameters for the different two-dimensional ferromagnetic semiconductors hosting large-gap QAH phase. $a, b, \lambda, \epsilon$ are the model parameters defined in Eqns.~(\ref{Eq_tb}) and (\ref{Eq_soc}) in unit of meV. $\Delta$ is the size of the topological gap. }
\label{Tab1}
\end{table}

{\it Antiferromagnetic QSH.} -- 
Following the standard strategy, by imposing ${\cal T}$ one can easily get a QSH model from superposing two copies of QAH insulators with opposite local magnetic moments. 
\begin{equation} \label{Eq:QSH}
H_{qsh}(k) = 
\begin{pmatrix}
H_{qah}(k) & 0 \cr
0 & {\cal T}H_{qah}{\cal T}^{-1}
\end{pmatrix}
\end{equation}
It is easy to understand that, if one QAH insulator is fully polarized along $z$-direction, the other copy will polarize along $-z$-direction, which would correspond to antiferromagnetic stacking of two layers of QAH systems. 
As the topological character of QSH system is fully determined by the topology of each QAH subsystem, the generic phase diagram for the QSH Hamiltonian Eq.~\ref{Eq:QSH} is exactly same as Fig.~\ref{Fig1} except for that the Chern number now corresponds to the spin Chern number defined as $c_{s}=\frac{1}{2}(c_{\uparrow}-c_{\downarrow})$, where $c_{\uparrow}$ and $c_{\downarrow}$ are the Chern numbers for the two spin-polarized QAH subsystems. 

So far all the experimentally confirmed QSH insulators, including HgTe/CdTe quantum well and bismuthene etc., are nonmagnetic. 
There is no local magnetic moment in these systems. 
Thus, it is not possible to identify the QAH component inside these QSH materials and, subsequently, to isolate them to get a QAH insulator. 
Actually, it is sufficient for a system to be a QSH insulator if it satisfies the time-reversal invariant QSH Hamiltonian Eq.~(\ref{Eq:QSH}) and, at the same time, $H_{qah}(k)$ carries nontrivial berry curvature. 
There is no need for $H_{qah}(k)$ to have a net local magnetic moment. 
HgTe/CdTe quantum well and bismuthene belong to this type, and they are known as nonmagnetic QSH insulators. 
Here, we construct a QSH system with local magnetic moment, which is qualitatively different from the other mentioned QSH systems. 
The local magnetic moment does not cancel each other, but the overall system respects the time-reversal symmetry.
It shall be named as antiferromagnetic QSH state, and it is feasible to isolate and recover the two copies of QAH systems. 

\begin{figure}[htbp]
\centering
\includegraphics[width=\linewidth]{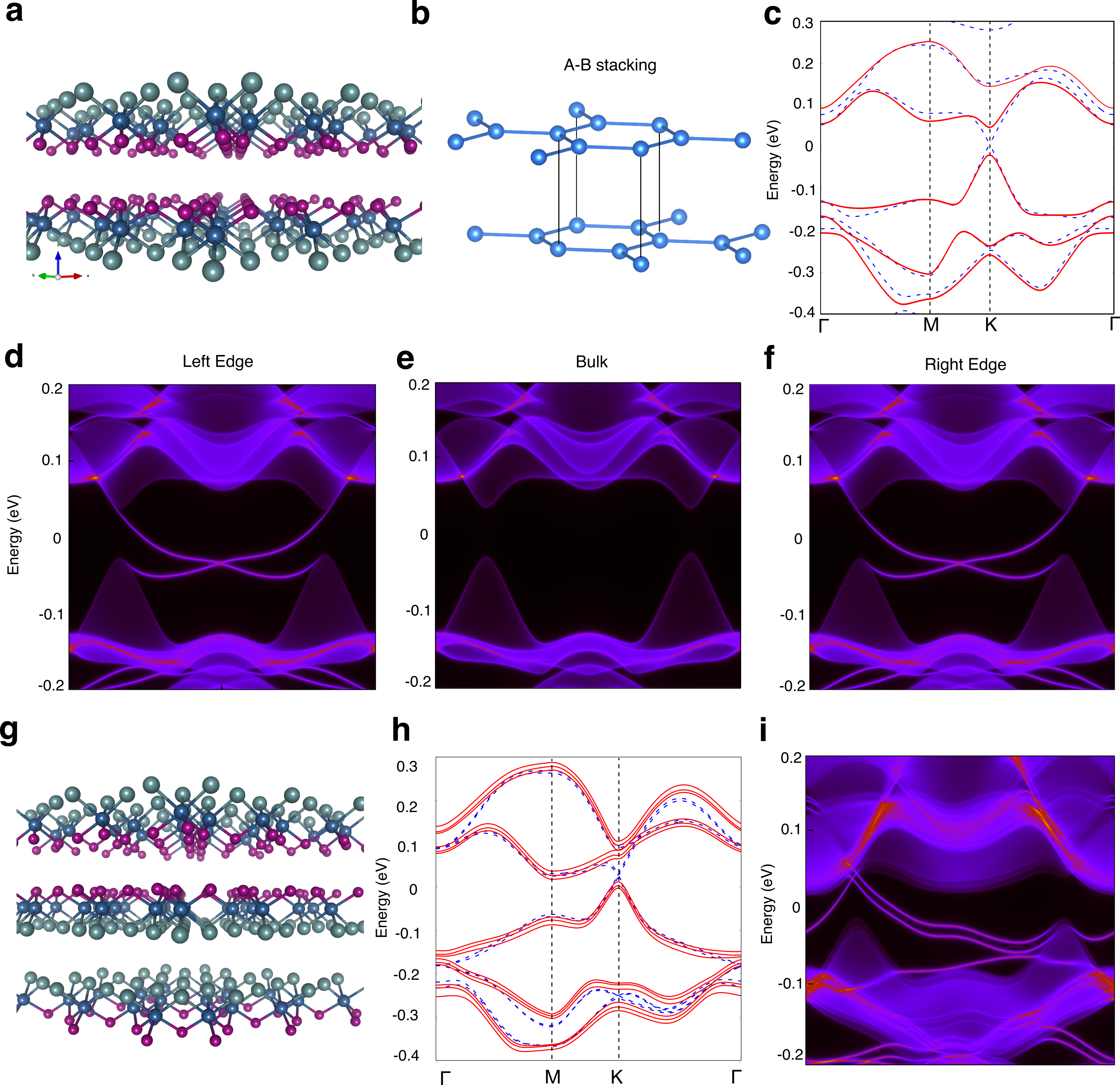}
\caption{QSH insulator composed from bilayer stacking of two copies of QAH insulators. (a) The side view of the structure model of bilayer Fe$_{2}$Cl$_{3}$I$_{3}$. (b) The two layers stack in A-B form, e.g. A sites of the top layer lay exactly on top of the B sites of the bottom layer. The B sites of the top layer and the A sites of the bottom layer stay in the middle the hexagon of the other layer. (c) The electronic structure of bilayer of Fe$_{2}$Cl$_{3}$I$_{3}$. The red solid line and the blue dashed line denote the bands with and without SOC. (d - f) The topological edge states along the zigzag edge. Compared to the bulk states, along each edge there exist two counter propagating edge modes charactering the QSH nature of the bilayer system. (g - i) The crystal structure, bulk electronic structure and the topological edge mode of trilayer Fe$_{2}$Cl$_{3}$I$_{3}$.}
\label{Fig4}
\end{figure}

The structure of antiferromagnetic QSH insulator constructed with iron halogenide is same as bilayer graphene. 
As an example, we showed bilayer Fe$_{2}$Cl$_{3}$I$_{3}$ stacked in A-B form in Fig.~\ref{Fig4} (b). 
The two layers are connected by van der Waals force, i.e. no direct electron hoping between the two layers. 
After fully relaxing the crystal, we obtained the electronic structures with and without SOC in Fig.~\ref{Fig4}(c) displayed as red solid line and blue dashed line, respectively.
Compared to Fig.~\ref{Fig3}(c), the electronic structure of the monolayer and bilayer Fe$_{2}$Cl$_{3}$I$_{3}$ is very similar indicating a small interlayer coupling. 
Although the interlayer hopping is forbidden due to the opposite spin polarization of the two layers, we found that there remains exchange couplings between them, which slightly gaps the Dirac linear band crossing at $\vec{K}$ and $\vec{K}^{\prime}$ in the absence of SOC. 
With respect to the Hamiltonian Eq.~(\ref{Eq:QSH}), it means that nonzero elements enter the off-diagonal block which couple the two QAH insulators. 
While, this small exchange coupling does not qualitatively change the topology of any layer. 
By direct calculating $Z_{2}$ topological invariant, we confirm bilayer iron halogenide as an antiferromagnetic QSH insulator.
Along both the left and the right edges of a zigzag ribbon, there exist two counter-propagating edge modes as illustrated in Fig.~\ref{Fig4} (d-e).
The two modes carrying opposite spin polarizations and each corresponds to a QAH edge mode. 
Their appearance, thus, exactly follows the construction of QSH states with QAH insulators and unbiasedly validates the conceptual relationship of QSH and QAH states.

{\it Discussions.} --
The proposed QAH and QSH models have been demonstrated in monolayer and bilayer iron halogenide. 
Experimentally, FeBr$_{3}$, FeCl$_{3}$ and FeI$_{3}$ crystalize in van der Waals quasi two-dimensional form and their successful synthesis have been reported~\cite{C9CP05084A, PhysRev.127.714, FeBr3-1, doi:10.1021/ja01145a511}. 
Due to the weak interlayer coupling, it is highly feasible to cleave these materials to realize the promising QSH and QAH states discussed above. 
The thermal and dynamical stability of their monolayers were also confirmed theoretically by different groups~\cite{Haastrup_2018, cryst7050121}.
We note that the bulk electronic structures of these materials are different from those of their thin film, which seems to indicate a three-dimensional nature of these materials. 
Cleaving these materials may not seem to be straightforward.
We note that the change of the electronic structure stems from the change of magnetic configuration of these system. 
There exist a low-spin to high-spin phase transition in these systems with the increase of sample thickness. 
The ground states of the thin film and the bulk are low-spin and high-spin states of iron $d$-electrons, respectively. 
The increase of film thickness reduces the energy gap between $t_{2g}$ and $e_{g}$ states, with the latter lying below. 
Thus, a low-spin configuration of iron $d^{5}$ with $S_{\mbox{eff}}=1/2$ at the thin film limit will change to a high-spin configuration with $S_{\mbox{eff}}=5/2$ in the bulk.
The change of electronic structure is not due to a strong interlayer coupling, thus, we believe the mechanical cleavage can be straightforwardly applied. 
As long as the film thickness is smaller than certain critical value, the desired low-spin states and the QAH(QSH) states will naturally appear. 
In our calculations, the low-spin states of the 3 and 4 layers of Fe$_{2}$Cl$_{3}$I$_{3}$ can also be stabilized. 
In Fig.~\ref{Fig4}(g - i), the corresponding crystal structure, bulk electronic structure and the topological edge modes for a 3-layer Fe$_{2}$Cl$_{3}$I$_{3}$ are shown, which confirm it as a QAH insulator. 
The different topological phases in the same system with even/odd number of layers and their phase transitions were also observed recently in magnetic topological insulator MnBi$_{2}$Te$_{4}$~\cite{PhysRevLett.122.206401, Lieaaw5685, PhysRevLett.122.107202, 2019arXiv190411468D, 2019arXiv190709947G, axion}. 
The axion insulator in even-number septuple-layer (SL) MnBi$_{2}$Te$_{4}$  transforms to the QAH insulator in odd-number SLs case.

{\it Conclusion.} --
 In this work, we propose a generic tight-binding model consisting of two correlated orbitals $d_{xy}/d_{x^{2}-y^{2}}$ with large local SOC strength. 
This model displays a rich topological phase diagram with three distinct topological states. 
One of these topological phases was shown to realize in a class of two-dimensional ferromagnetic insulators, which all display a large-gap QAH state featuring a possible room-temperature spintronic application. 
By layer-stacking, these QAH material candidates were further shown to assemble stable QSH insulators with non-vanishing local magnetic moment, i.e. the long-thought antiferromagnetic QSH insulator.   
The high mechanical adjustability makes these system perfect material platform for switching between QSH and QAH states.
 
 {\it Acknowledgement} --
This work was supported by the National Key RD Program of China (2017YFE0131300).
G. Li would like to thank the financial support from the starting grant of ShanghaiTech University and the Program for Professor of Special Appointment (Shanghai Eastern Scholar).
Calculations were carried out at the HPC Platform of ShanghaiTech University Library and Information Services, and at School of Physical Science and Technology.




\bibliographystyle{apsrev4-1}
\bibliography{ref}  

\begin{thebibliography}{28}%
\makeatletter
\providecommand \@ifxundefined [1]{%
 \@ifx{#1\undefined}
}%
\providecommand \@ifnum [1]{%
 \ifnum #1\expandafter \@firstoftwo
 \else \expandafter \@secondoftwo
 \fi
}%
\providecommand \@ifx [1]{%
 \ifx #1\expandafter \@firstoftwo
 \else \expandafter \@secondoftwo
 \fi
}%
\providecommand \natexlab [1]{#1}%
\providecommand \enquote  [1]{``#1''}%
\providecommand \bibnamefont  [1]{#1}%
\providecommand \bibfnamefont [1]{#1}%
\providecommand \citenamefont [1]{#1}%
\providecommand \href@noop [0]{\@secondoftwo}%
\providecommand \href [0]{\begingroup \@sanitize@url \@href}%
\providecommand \@href[1]{\@@startlink{#1}\@@href}%
\providecommand \@@href[1]{\endgroup#1\@@endlink}%
\providecommand \@sanitize@url [0]{\catcode `\\12\catcode `\$12\catcode
  `\&12\catcode `\#12\catcode `\^12\catcode `\_12\catcode `\%12\relax}%
\providecommand \@@startlink[1]{}%
\providecommand \@@endlink[0]{}%
\providecommand \url  [0]{\begingroup\@sanitize@url \@url }%
\providecommand \@url [1]{\endgroup\@href {#1}{\urlprefix }}%
\providecommand \urlprefix  [0]{URL }%
\providecommand \Eprint [0]{\href }%
\providecommand \doibase [0]{http://dx.doi.org/}%
\providecommand \selectlanguage [0]{\@gobble}%
\providecommand \bibinfo  [0]{\@secondoftwo}%
\providecommand \bibfield  [0]{\@secondoftwo}%
\providecommand \translation [1]{[#1]}%
\providecommand \BibitemOpen [0]{}%
\providecommand \bibitemStop [0]{}%
\providecommand \bibitemNoStop [0]{.\EOS\space}%
\providecommand \EOS [0]{\spacefactor3000\relax}%
\providecommand \BibitemShut  [1]{\csname bibitem#1\endcsname}%
\let\auto@bib@innerbib\@empty
\bibitem [{\citenamefont {Klitzing}\ \emph {et~al.}(1980)\citenamefont
  {Klitzing}, \citenamefont {Dorda},\ and\ \citenamefont
  {Pepper}}]{PhysRevLett.45.494}%
  \BibitemOpen
  \bibfield  {author} {\bibinfo {author} {\bibfnamefont {K.~v.}\ \bibnamefont
  {Klitzing}}, \bibinfo {author} {\bibfnamefont {G.}~\bibnamefont {Dorda}}, \
  and\ \bibinfo {author} {\bibfnamefont {M.}~\bibnamefont {Pepper}},\ }\href
  {\doibase 10.1103/PhysRevLett.45.494} {\bibfield  {journal} {\bibinfo
  {journal} {Phys. Rev. Lett.}\ }\textbf {\bibinfo {volume} {45}},\ \bibinfo
  {pages} {494} (\bibinfo {year} {1980})}\BibitemShut {NoStop}%
\bibitem [{\citenamefont {Hasan}\ and\ \citenamefont
  {Kane}(2010)}]{RevModPhys.82.3045}%
  \BibitemOpen
  \bibfield  {author} {\bibinfo {author} {\bibfnamefont {M.~Z.}\ \bibnamefont
  {Hasan}}\ and\ \bibinfo {author} {\bibfnamefont {C.~L.}\ \bibnamefont
  {Kane}},\ }\href {\doibase 10.1103/RevModPhys.82.3045} {\bibfield  {journal}
  {\bibinfo  {journal} {Rev. Mod. Phys.}\ }\textbf {\bibinfo {volume} {82}},\
  \bibinfo {pages} {3045} (\bibinfo {year} {2010})}\BibitemShut {NoStop}%
\bibitem [{\citenamefont {Qi}\ and\ \citenamefont
  {Zhang}(2011)}]{RevModPhys.83.1057}%
  \BibitemOpen
  \bibfield  {author} {\bibinfo {author} {\bibfnamefont {X.-L.}\ \bibnamefont
  {Qi}}\ and\ \bibinfo {author} {\bibfnamefont {S.-C.}\ \bibnamefont {Zhang}},\
  }\href {\doibase 10.1103/RevModPhys.83.1057} {\bibfield  {journal} {\bibinfo
  {journal} {Rev. Mod. Phys.}\ }\textbf {\bibinfo {volume} {83}},\ \bibinfo
  {pages} {1057} (\bibinfo {year} {2011})}\BibitemShut {NoStop}%
\bibitem [{\citenamefont {Thouless}\ \emph {et~al.}(1982)\citenamefont
  {Thouless}, \citenamefont {Kohmoto}, \citenamefont {Nightingale},\ and\
  \citenamefont {den Nijs}}]{PhysRevLett.49.405}%
  \BibitemOpen
  \bibfield  {author} {\bibinfo {author} {\bibfnamefont {D.~J.}\ \bibnamefont
  {Thouless}}, \bibinfo {author} {\bibfnamefont {M.}~\bibnamefont {Kohmoto}},
  \bibinfo {author} {\bibfnamefont {M.~P.}\ \bibnamefont {Nightingale}}, \ and\
  \bibinfo {author} {\bibfnamefont {M.}~\bibnamefont {den Nijs}},\ }\href
  {\doibase 10.1103/PhysRevLett.49.405} {\bibfield  {journal} {\bibinfo
  {journal} {Phys. Rev. Lett.}\ }\textbf {\bibinfo {volume} {49}},\ \bibinfo
  {pages} {405} (\bibinfo {year} {1982})}\BibitemShut {NoStop}%
\bibitem [{\citenamefont {Berry}(1984)}]{Berry1984}%
  \BibitemOpen
  \bibfield  {author} {\bibinfo {author} {\bibfnamefont {M.~V.}\ \bibnamefont
  {Berry}},\ }\href {\doibase 10.1098/rspa.1984.0023} {\bibfield  {journal}
  {\bibinfo  {journal} {Proceedings of the Royal Society of London. A.
  Mathematical and Physical Sciences}\ }\textbf {\bibinfo {volume} {392}},\
  \bibinfo {pages} {45} (\bibinfo {year} {1984})}\BibitemShut {NoStop}%
\bibitem [{\citenamefont {Haldane}(1988)}]{PhysRevLett.61.2015}%
  \BibitemOpen
  \bibfield  {author} {\bibinfo {author} {\bibfnamefont {F.~D.~M.}\
  \bibnamefont {Haldane}},\ }\href {\doibase 10.1103/PhysRevLett.61.2015}
  {\bibfield  {journal} {\bibinfo  {journal} {Phys. Rev. Lett.}\ }\textbf
  {\bibinfo {volume} {61}},\ \bibinfo {pages} {2015} (\bibinfo {year}
  {1988})}\BibitemShut {NoStop}%
\bibitem [{\citenamefont {Kane}\ and\ \citenamefont {Mele}(2005)}]{Kane2005}%
  \BibitemOpen
  \bibfield  {author} {\bibinfo {author} {\bibfnamefont {C.~L.}\ \bibnamefont
  {Kane}}\ and\ \bibinfo {author} {\bibfnamefont {E.~J.}\ \bibnamefont
  {Mele}},\ }\href@noop {} {\bibfield  {journal} {\bibinfo  {journal} {Phys.
  Rev. Lett.}\ }\textbf {\bibinfo {volume} {95}} (\bibinfo {year}
  {2005})}\BibitemShut {NoStop}%
\bibitem [{\citenamefont {Bernevig}\ \emph {et~al.}(2006)\citenamefont
  {Bernevig}, \citenamefont {Hughes},\ and\ \citenamefont {Zhang}}]{hgte1}%
  \BibitemOpen
  \bibfield  {author} {\bibinfo {author} {\bibfnamefont {B.~A.}\ \bibnamefont
  {Bernevig}}, \bibinfo {author} {\bibfnamefont {T.~L.}\ \bibnamefont
  {Hughes}}, \ and\ \bibinfo {author} {\bibfnamefont {S.-C.}\ \bibnamefont
  {Zhang}},\ }\href@noop {} {\bibfield  {journal} {\bibinfo  {journal}
  {Science}\ }\textbf {\bibinfo {volume} {314}},\ \bibinfo {pages} {1757}
  (\bibinfo {year} {2006})}\BibitemShut {NoStop}%
\bibitem [{\citenamefont {König}\ \emph {et~al.}(2007)\citenamefont {König},
  \citenamefont {Wiedmann}, \citenamefont {Brüne}, \citenamefont {Roth},
  \citenamefont {Buhmann}, \citenamefont {Molenkamp}, \citenamefont {Qi},\ and\
  \citenamefont {Zhang}}]{hgte2}%
  \BibitemOpen
  \bibfield  {author} {\bibinfo {author} {\bibfnamefont {M.}~\bibnamefont
  {König}}, \bibinfo {author} {\bibfnamefont {S.}~\bibnamefont {Wiedmann}},
  \bibinfo {author} {\bibfnamefont {C.}~\bibnamefont {Brüne}}, \bibinfo
  {author} {\bibfnamefont {A.}~\bibnamefont {Roth}}, \bibinfo {author}
  {\bibfnamefont {H.}~\bibnamefont {Buhmann}}, \bibinfo {author} {\bibfnamefont
  {L.~W.}\ \bibnamefont {Molenkamp}}, \bibinfo {author} {\bibfnamefont {X.-L.}\
  \bibnamefont {Qi}}, \ and\ \bibinfo {author} {\bibfnamefont {S.-C.}\
  \bibnamefont {Zhang}},\ }\href@noop {} {\bibfield  {journal} {\bibinfo
  {journal} {Science}\ }\textbf {\bibinfo {volume} {318}},\ \bibinfo {pages}
  {766} (\bibinfo {year} {2007})}\BibitemShut {NoStop}%
\bibitem [{\citenamefont {Reis}\ \emph {et~al.}(2017)\citenamefont {Reis},
  \citenamefont {Li}, \citenamefont {Dudy}, \citenamefont {Bauernfeind},
  \citenamefont {Glass}, \citenamefont {Hanke}, \citenamefont {Thomale},
  \citenamefont {Sch{\"a}fer},\ and\ \citenamefont {Claessen}}]{Reis287}%
  \BibitemOpen
  \bibfield  {author} {\bibinfo {author} {\bibfnamefont {F.}~\bibnamefont
  {Reis}}, \bibinfo {author} {\bibfnamefont {G.}~\bibnamefont {Li}}, \bibinfo
  {author} {\bibfnamefont {L.}~\bibnamefont {Dudy}}, \bibinfo {author}
  {\bibfnamefont {M.}~\bibnamefont {Bauernfeind}}, \bibinfo {author}
  {\bibfnamefont {S.}~\bibnamefont {Glass}}, \bibinfo {author} {\bibfnamefont
  {W.}~\bibnamefont {Hanke}}, \bibinfo {author} {\bibfnamefont
  {R.}~\bibnamefont {Thomale}}, \bibinfo {author} {\bibfnamefont
  {J.}~\bibnamefont {Sch{\"a}fer}}, \ and\ \bibinfo {author} {\bibfnamefont
  {R.}~\bibnamefont {Claessen}},\ }\href {\doibase 10.1126/science.aai8142}
  {\bibfield  {journal} {\bibinfo  {journal} {Science}\ }\textbf {\bibinfo
  {volume} {357}},\ \bibinfo {pages} {287} (\bibinfo {year}
  {2017})}\BibitemShut {NoStop}%
\bibitem [{\citenamefont {Li}\ \emph {et~al.}(2018)\citenamefont {Li},
  \citenamefont {Hanke}, \citenamefont {Hankiewicz}, \citenamefont {Reis},
  \citenamefont {Sch\"afer}, \citenamefont {Claessen}, \citenamefont {Wu},\
  and\ \citenamefont {Thomale}}]{PhysRevB.98.165146}%
  \BibitemOpen
  \bibfield  {author} {\bibinfo {author} {\bibfnamefont {G.}~\bibnamefont
  {Li}}, \bibinfo {author} {\bibfnamefont {W.}~\bibnamefont {Hanke}}, \bibinfo
  {author} {\bibfnamefont {E.~M.}\ \bibnamefont {Hankiewicz}}, \bibinfo
  {author} {\bibfnamefont {F.}~\bibnamefont {Reis}}, \bibinfo {author}
  {\bibfnamefont {J.}~\bibnamefont {Sch\"afer}}, \bibinfo {author}
  {\bibfnamefont {R.}~\bibnamefont {Claessen}}, \bibinfo {author}
  {\bibfnamefont {C.}~\bibnamefont {Wu}}, \ and\ \bibinfo {author}
  {\bibfnamefont {R.}~\bibnamefont {Thomale}},\ }\href {\doibase
  10.1103/PhysRevB.98.165146} {\bibfield  {journal} {\bibinfo  {journal} {Phys.
  Rev. B}\ }\textbf {\bibinfo {volume} {98}},\ \bibinfo {pages} {165146}
  (\bibinfo {year} {2018})}\BibitemShut {NoStop}%
\bibitem [{\citenamefont {Dominguez}\ \emph {et~al.}(2018)\citenamefont
  {Dominguez}, \citenamefont {Scharf}, \citenamefont {Li}, \citenamefont
  {Sch\"afer}, \citenamefont {Claessen}, \citenamefont {Hanke}, \citenamefont
  {Thomale},\ and\ \citenamefont {Hankiewicz}}]{PhysRevB.98.161407}%
  \BibitemOpen
  \bibfield  {author} {\bibinfo {author} {\bibfnamefont {F.}~\bibnamefont
  {Dominguez}}, \bibinfo {author} {\bibfnamefont {B.}~\bibnamefont {Scharf}},
  \bibinfo {author} {\bibfnamefont {G.}~\bibnamefont {Li}}, \bibinfo {author}
  {\bibfnamefont {J.}~\bibnamefont {Sch\"afer}}, \bibinfo {author}
  {\bibfnamefont {R.}~\bibnamefont {Claessen}}, \bibinfo {author}
  {\bibfnamefont {W.}~\bibnamefont {Hanke}}, \bibinfo {author} {\bibfnamefont
  {R.}~\bibnamefont {Thomale}}, \ and\ \bibinfo {author} {\bibfnamefont
  {E.~M.}\ \bibnamefont {Hankiewicz}},\ }\href {\doibase
  10.1103/PhysRevB.98.161407} {\bibfield  {journal} {\bibinfo  {journal} {Phys.
  Rev. B}\ }\textbf {\bibinfo {volume} {98}},\ \bibinfo {pages} {161407}
  (\bibinfo {year} {2018})}\BibitemShut {NoStop}%
\bibitem [{\citenamefont {Liu}\ \emph {et~al.}(2009)\citenamefont {Liu},
  \citenamefont {Liu}, \citenamefont {Xu}, \citenamefont {Qi},\ and\
  \citenamefont {Zhang}}]{PhysRevLett.102.156603}%
  \BibitemOpen
  \bibfield  {author} {\bibinfo {author} {\bibfnamefont {Q.}~\bibnamefont
  {Liu}}, \bibinfo {author} {\bibfnamefont {C.-X.}\ \bibnamefont {Liu}},
  \bibinfo {author} {\bibfnamefont {C.}~\bibnamefont {Xu}}, \bibinfo {author}
  {\bibfnamefont {X.-L.}\ \bibnamefont {Qi}}, \ and\ \bibinfo {author}
  {\bibfnamefont {S.-C.}\ \bibnamefont {Zhang}},\ }\href {\doibase
  10.1103/PhysRevLett.102.156603} {\bibfield  {journal} {\bibinfo  {journal}
  {Phys. Rev. Lett.}\ }\textbf {\bibinfo {volume} {102}},\ \bibinfo {pages}
  {156603} (\bibinfo {year} {2009})}\BibitemShut {NoStop}%
\bibitem [{\citenamefont {Yu}\ \emph {et~al.}(2010)\citenamefont {Yu},
  \citenamefont {Zhang}, \citenamefont {Zhang}, \citenamefont {Zhang},
  \citenamefont {Dai},\ and\ \citenamefont {Fang}}]{Yu61}%
  \BibitemOpen
  \bibfield  {author} {\bibinfo {author} {\bibfnamefont {R.}~\bibnamefont
  {Yu}}, \bibinfo {author} {\bibfnamefont {W.}~\bibnamefont {Zhang}}, \bibinfo
  {author} {\bibfnamefont {H.-J.}\ \bibnamefont {Zhang}}, \bibinfo {author}
  {\bibfnamefont {S.-C.}\ \bibnamefont {Zhang}}, \bibinfo {author}
  {\bibfnamefont {X.}~\bibnamefont {Dai}}, \ and\ \bibinfo {author}
  {\bibfnamefont {Z.}~\bibnamefont {Fang}},\ }\href {\doibase
  10.1126/science.1187485} {\bibfield  {journal} {\bibinfo  {journal}
  {Science}\ }\textbf {\bibinfo {volume} {329}},\ \bibinfo {pages} {61}
  (\bibinfo {year} {2010})}\BibitemShut {NoStop}%
\bibitem [{\citenamefont {Slater}\ and\ \citenamefont
  {Koster}(1954)}]{PhysRev.94.1498}%
  \BibitemOpen
  \bibfield  {author} {\bibinfo {author} {\bibfnamefont {J.~C.}\ \bibnamefont
  {Slater}}\ and\ \bibinfo {author} {\bibfnamefont {G.~F.}\ \bibnamefont
  {Koster}},\ }\href {\doibase 10.1103/PhysRev.94.1498} {\bibfield  {journal}
  {\bibinfo  {journal} {Phys. Rev.}\ }\textbf {\bibinfo {volume} {94}},\
  \bibinfo {pages} {1498} (\bibinfo {year} {1954})}\BibitemShut {NoStop}%
\bibitem [{\citenamefont {Zhang}\ and\ \citenamefont
  {Liu}(2017)}]{zhang2017intrinsic}%
  \BibitemOpen
  \bibfield  {author} {\bibinfo {author} {\bibfnamefont {S.-H.}\ \bibnamefont
  {Zhang}}\ and\ \bibinfo {author} {\bibfnamefont {B.-G.}\ \bibnamefont
  {Liu}},\ }\href@noop {} {\enquote {\bibinfo {title} {Intrinsic 2d
  ferromagnetism, quantum anomalous hall conductivity, and fully-spin-polarized
  edge states of febr3 monolayer},}\ } (\bibinfo {year} {2017}),\ \Eprint
  {http://arxiv.org/abs/1706.08943} {arXiv:1706.08943 [cond-mat.mes-hall]}
  \BibitemShut {NoStop}%
\bibitem [{\citenamefont {Sun}\ \emph {et~al.}(2020)\citenamefont {Sun},
  \citenamefont {Zhong}, \citenamefont {Cui}, \citenamefont {Shi},
  \citenamefont {Hao}, \citenamefont {Xu},\ and\ \citenamefont
  {Li}}]{C9CP05084A}%
  \BibitemOpen
  \bibfield  {author} {\bibinfo {author} {\bibfnamefont {J.}~\bibnamefont
  {Sun}}, \bibinfo {author} {\bibfnamefont {X.}~\bibnamefont {Zhong}}, \bibinfo
  {author} {\bibfnamefont {W.}~\bibnamefont {Cui}}, \bibinfo {author}
  {\bibfnamefont {J.}~\bibnamefont {Shi}}, \bibinfo {author} {\bibfnamefont
  {J.}~\bibnamefont {Hao}}, \bibinfo {author} {\bibfnamefont {M.}~\bibnamefont
  {Xu}}, \ and\ \bibinfo {author} {\bibfnamefont {Y.}~\bibnamefont {Li}},\
  }\href {\doibase 10.1107/S0021889888013913} {\bibfield  {journal} {\bibinfo
  {journal} {Phys. Chem. Chem. Phys.}\ }\textbf {\bibinfo {volume} {22}},\
  (\bibinfo {year} {2020})}\BibitemShut {NoStop}%
\bibitem [{\citenamefont {Cable}\ \emph {et~al.}(1962)\citenamefont {Cable},
  \citenamefont {Wilkinson}, \citenamefont {Wollan},\ and\ \citenamefont
  {Koehler}}]{PhysRev.127.714}%
  \BibitemOpen
  \bibfield  {author} {\bibinfo {author} {\bibfnamefont {J.~W.}\ \bibnamefont
  {Cable}}, \bibinfo {author} {\bibfnamefont {M.~K.}\ \bibnamefont
  {Wilkinson}}, \bibinfo {author} {\bibfnamefont {E.~O.}\ \bibnamefont
  {Wollan}}, \ and\ \bibinfo {author} {\bibfnamefont {W.~C.}\ \bibnamefont
  {Koehler}},\ }\href {\doibase 10.1103/PhysRev.127.714} {\bibfield  {journal}
  {\bibinfo  {journal} {Phys. Rev.}\ }\textbf {\bibinfo {volume} {127}},\
  \bibinfo {pages} {714} (\bibinfo {year} {1962})}\BibitemShut {NoStop}%
\bibitem [{\citenamefont {Armbruster}\ \emph {et~al.}(2000)\citenamefont
  {Armbruster}, \citenamefont {Ludwig}, \citenamefont {Rotter}, \citenamefont
  {Thiele},\ and\ \citenamefont {Oppermann}}]{FeBr3-1}%
  \BibitemOpen
  \bibfield  {author} {\bibinfo {author} {\bibfnamefont {M.}~\bibnamefont
  {Armbruster}}, \bibinfo {author} {\bibfnamefont {T.}~\bibnamefont {Ludwig}},
  \bibinfo {author} {\bibfnamefont {H.~W.}\ \bibnamefont {Rotter}}, \bibinfo
  {author} {\bibfnamefont {G.}~\bibnamefont {Thiele}}, \ and\ \bibinfo {author}
  {\bibfnamefont {H.}~\bibnamefont {Oppermann}},\ }\href {\doibase
  10.1002/(SICI)1521-3749(200001)626:1<187::AID-ZAAC187>3.0.CO;2-X} {\bibfield
  {journal} {\bibinfo  {journal} {Zeitschrift für anorganische und allgemeine
  Chemie}\ }\textbf {\bibinfo {volume} {626}},\ \bibinfo {pages} {187}
  (\bibinfo {year} {2000})}\BibitemShut {NoStop}%
\bibitem [{\citenamefont {Gregory}(1951)}]{doi:10.1021/ja01145a511}%
  \BibitemOpen
  \bibfield  {author} {\bibinfo {author} {\bibfnamefont {N.~W.}\ \bibnamefont
  {Gregory}},\ }\href {\doibase 10.1021/ja01145a511} {\bibfield  {journal}
  {\bibinfo  {journal} {Journal of the American Chemical Society}\ }\textbf
  {\bibinfo {volume} {73}},\ \bibinfo {pages} {472} (\bibinfo {year}
  {1951})}\BibitemShut {NoStop}%
\bibitem [{\citenamefont {Haastrup}\ \emph {et~al.}(2018)\citenamefont
  {Haastrup}, \citenamefont {Strange}, \citenamefont {Pandey}, \citenamefont
  {Deilmann}, \citenamefont {Schmidt}, \citenamefont {Hinsche}, \citenamefont
  {Gjerding}, \citenamefont {Torelli}, \citenamefont {Larsen}, \citenamefont
  {Riis-Jensen}, \citenamefont {Gath}, \citenamefont {Jacobsen}, \citenamefont
  {Mortensen}, \citenamefont {Olsen},\ and\ \citenamefont
  {Thygesen}}]{Haastrup_2018}%
  \BibitemOpen
  \bibfield  {author} {\bibinfo {author} {\bibfnamefont {S.}~\bibnamefont
  {Haastrup}}, \bibinfo {author} {\bibfnamefont {M.}~\bibnamefont {Strange}},
  \bibinfo {author} {\bibfnamefont {M.}~\bibnamefont {Pandey}}, \bibinfo
  {author} {\bibfnamefont {T.}~\bibnamefont {Deilmann}}, \bibinfo {author}
  {\bibfnamefont {P.~S.}\ \bibnamefont {Schmidt}}, \bibinfo {author}
  {\bibfnamefont {N.~F.}\ \bibnamefont {Hinsche}}, \bibinfo {author}
  {\bibfnamefont {M.~N.}\ \bibnamefont {Gjerding}}, \bibinfo {author}
  {\bibfnamefont {D.}~\bibnamefont {Torelli}}, \bibinfo {author} {\bibfnamefont
  {P.~M.}\ \bibnamefont {Larsen}}, \bibinfo {author} {\bibfnamefont {A.~C.}\
  \bibnamefont {Riis-Jensen}}, \bibinfo {author} {\bibfnamefont
  {J.}~\bibnamefont {Gath}}, \bibinfo {author} {\bibfnamefont {K.~W.}\
  \bibnamefont {Jacobsen}}, \bibinfo {author} {\bibfnamefont {J.~J.}\
  \bibnamefont {Mortensen}}, \bibinfo {author} {\bibfnamefont {T.}~\bibnamefont
  {Olsen}}, \ and\ \bibinfo {author} {\bibfnamefont {K.~S.}\ \bibnamefont
  {Thygesen}},\ }\href {\doibase 10.1088/2053-1583/aacfc1} {\bibfield
  {journal} {\bibinfo  {journal} {2D Materials}\ }\textbf {\bibinfo {volume}
  {5}},\ \bibinfo {pages} {042002} (\bibinfo {year} {2018})}\BibitemShut
  {NoStop}%
\bibitem [{\citenamefont {McGuire}(2017)}]{cryst7050121}%
  \BibitemOpen
  \bibfield  {author} {\bibinfo {author} {\bibfnamefont {M.~A.}\ \bibnamefont
  {McGuire}},\ }\href {\doibase 10.3390/cryst7050121} {\bibfield  {journal}
  {\bibinfo  {journal} {Crystals}\ }\textbf {\bibinfo {volume} {7}} (\bibinfo
  {year} {2017}),\ 10.3390/cryst7050121}\BibitemShut {NoStop}%
\bibitem [{\citenamefont {Zhang}\ \emph {et~al.}(2019)\citenamefont {Zhang},
  \citenamefont {Shi}, \citenamefont {Zhu}, \citenamefont {Xing}, \citenamefont
  {Zhang},\ and\ \citenamefont {Wang}}]{PhysRevLett.122.206401}%
  \BibitemOpen
  \bibfield  {author} {\bibinfo {author} {\bibfnamefont {D.}~\bibnamefont
  {Zhang}}, \bibinfo {author} {\bibfnamefont {M.}~\bibnamefont {Shi}}, \bibinfo
  {author} {\bibfnamefont {T.}~\bibnamefont {Zhu}}, \bibinfo {author}
  {\bibfnamefont {D.}~\bibnamefont {Xing}}, \bibinfo {author} {\bibfnamefont
  {H.}~\bibnamefont {Zhang}}, \ and\ \bibinfo {author} {\bibfnamefont
  {J.}~\bibnamefont {Wang}},\ }\href {\doibase 10.1103/PhysRevLett.122.206401}
  {\bibfield  {journal} {\bibinfo  {journal} {Phys. Rev. Lett.}\ }\textbf
  {\bibinfo {volume} {122}},\ \bibinfo {pages} {206401} (\bibinfo {year}
  {2019})}\BibitemShut {NoStop}%
\bibitem [{\citenamefont {Li}\ \emph {et~al.}(2019)\citenamefont {Li},
  \citenamefont {Li}, \citenamefont {Du}, \citenamefont {Wang}, \citenamefont
  {Gu}, \citenamefont {Zhang}, \citenamefont {He}, \citenamefont {Duan},\ and\
  \citenamefont {Xu}}]{Lieaaw5685}%
  \BibitemOpen
  \bibfield  {author} {\bibinfo {author} {\bibfnamefont {J.}~\bibnamefont
  {Li}}, \bibinfo {author} {\bibfnamefont {Y.}~\bibnamefont {Li}}, \bibinfo
  {author} {\bibfnamefont {S.}~\bibnamefont {Du}}, \bibinfo {author}
  {\bibfnamefont {Z.}~\bibnamefont {Wang}}, \bibinfo {author} {\bibfnamefont
  {B.-L.}\ \bibnamefont {Gu}}, \bibinfo {author} {\bibfnamefont {S.-C.}\
  \bibnamefont {Zhang}}, \bibinfo {author} {\bibfnamefont {K.}~\bibnamefont
  {He}}, \bibinfo {author} {\bibfnamefont {W.}~\bibnamefont {Duan}}, \ and\
  \bibinfo {author} {\bibfnamefont {Y.}~\bibnamefont {Xu}},\ }\href {\doibase
  10.1126/sciadv.aaw5685} {\bibfield  {journal} {\bibinfo  {journal} {Science
  Advances}\ }\textbf {\bibinfo {volume} {5}} (\bibinfo {year} {2019}),\
  10.1126/sciadv.aaw5685}\BibitemShut {NoStop}%
\bibitem [{\citenamefont {Otrokov}\ \emph {et~al.}(2019)\citenamefont
  {Otrokov}, \citenamefont {Rusinov}, \citenamefont {Blanco-Rey}, \citenamefont
  {Hoffmann}, \citenamefont {Vyazovskaya}, \citenamefont {Eremeev},
  \citenamefont {Ernst}, \citenamefont {Echenique}, \citenamefont {Arnau},\
  and\ \citenamefont {Chulkov}}]{PhysRevLett.122.107202}%
  \BibitemOpen
  \bibfield  {author} {\bibinfo {author} {\bibfnamefont {M.~M.}\ \bibnamefont
  {Otrokov}}, \bibinfo {author} {\bibfnamefont {I.~P.}\ \bibnamefont
  {Rusinov}}, \bibinfo {author} {\bibfnamefont {M.}~\bibnamefont {Blanco-Rey}},
  \bibinfo {author} {\bibfnamefont {M.}~\bibnamefont {Hoffmann}}, \bibinfo
  {author} {\bibfnamefont {A.~Y.}\ \bibnamefont {Vyazovskaya}}, \bibinfo
  {author} {\bibfnamefont {S.~V.}\ \bibnamefont {Eremeev}}, \bibinfo {author}
  {\bibfnamefont {A.}~\bibnamefont {Ernst}}, \bibinfo {author} {\bibfnamefont
  {P.~M.}\ \bibnamefont {Echenique}}, \bibinfo {author} {\bibfnamefont
  {A.}~\bibnamefont {Arnau}}, \ and\ \bibinfo {author} {\bibfnamefont {E.~V.}\
  \bibnamefont {Chulkov}},\ }\href {\doibase 10.1103/PhysRevLett.122.107202}
  {\bibfield  {journal} {\bibinfo  {journal} {Phys. Rev. Lett.}\ }\textbf
  {\bibinfo {volume} {122}},\ \bibinfo {pages} {107202} (\bibinfo {year}
  {2019})}\BibitemShut {NoStop}%
\bibitem [{\citenamefont {{Deng}}\ \emph {et~al.}(2019)\citenamefont {{Deng}},
  \citenamefont {{Yu}}, \citenamefont {{Zhu Shi}}, \citenamefont {{Wang}},
  \citenamefont {{Chen}},\ and\ \citenamefont {{Zhang}}}]{2019arXiv190411468D}%
  \BibitemOpen
  \bibfield  {author} {\bibinfo {author} {\bibfnamefont {Y.}~\bibnamefont
  {{Deng}}}, \bibinfo {author} {\bibfnamefont {Y.}~\bibnamefont {{Yu}}},
  \bibinfo {author} {\bibfnamefont {M.}~\bibnamefont {{Zhu Shi}}}, \bibinfo
  {author} {\bibfnamefont {J.}~\bibnamefont {{Wang}}}, \bibinfo {author}
  {\bibfnamefont {X.~H.}\ \bibnamefont {{Chen}}}, \ and\ \bibinfo {author}
  {\bibfnamefont {Y.}~\bibnamefont {{Zhang}}},\ }\href@noop {} {\bibfield
  {journal} {\bibinfo  {journal} {arXiv e-prints}\ ,\ \bibinfo {eid}
  {arXiv:1904.11468}} (\bibinfo {year} {2019})},\ \Eprint
  {http://arxiv.org/abs/1904.11468} {arXiv:1904.11468 [cond-mat.mtrl-sci]}
  \BibitemShut {NoStop}%
\bibitem [{\citenamefont {{Ge}}\ \emph {et~al.}(2019)\citenamefont {{Ge}},
  \citenamefont {{Liu}}, \citenamefont {{Li}}, \citenamefont {{Li}},
  \citenamefont {{Luo}}, \citenamefont {{Wu}}, \citenamefont {{Xu}},\ and\
  \citenamefont {{Wang}}}]{2019arXiv190709947G}%
  \BibitemOpen
  \bibfield  {author} {\bibinfo {author} {\bibfnamefont {J.}~\bibnamefont
  {{Ge}}}, \bibinfo {author} {\bibfnamefont {Y.}~\bibnamefont {{Liu}}},
  \bibinfo {author} {\bibfnamefont {J.}~\bibnamefont {{Li}}}, \bibinfo {author}
  {\bibfnamefont {H.}~\bibnamefont {{Li}}}, \bibinfo {author} {\bibfnamefont
  {T.}~\bibnamefont {{Luo}}}, \bibinfo {author} {\bibfnamefont
  {Y.}~\bibnamefont {{Wu}}}, \bibinfo {author} {\bibfnamefont {Y.}~\bibnamefont
  {{Xu}}}, \ and\ \bibinfo {author} {\bibfnamefont {J.}~\bibnamefont
  {{Wang}}},\ }\href@noop {} {\bibfield  {journal} {\bibinfo  {journal} {arXiv
  e-prints}\ ,\ \bibinfo {eid} {arXiv:1907.09947}} (\bibinfo {year} {2019})},\
  \Eprint {http://arxiv.org/abs/1907.09947} {arXiv:1907.09947
  [cond-mat.mes-hall]} \BibitemShut {NoStop}%
\bibitem [{\citenamefont {Liu}\ \emph {et~al.}(2020)\citenamefont {Liu},
  \citenamefont {Wang}, \citenamefont {Li}, \citenamefont {Wu}, \citenamefont
  {Li}, \citenamefont {Li}, \citenamefont {He}, \citenamefont {Xu},
  \citenamefont {Zhang},\ and\ \citenamefont {Wang}}]{axion}%
  \BibitemOpen
  \bibfield  {author} {\bibinfo {author} {\bibfnamefont {C.}~\bibnamefont
  {Liu}}, \bibinfo {author} {\bibfnamefont {Y.}~\bibnamefont {Wang}}, \bibinfo
  {author} {\bibfnamefont {H.}~\bibnamefont {Li}}, \bibinfo {author}
  {\bibfnamefont {Y.}~\bibnamefont {Wu}}, \bibinfo {author} {\bibfnamefont
  {Y.}~\bibnamefont {Li}}, \bibinfo {author} {\bibfnamefont {J.}~\bibnamefont
  {Li}}, \bibinfo {author} {\bibfnamefont {K.}~\bibnamefont {He}}, \bibinfo
  {author} {\bibfnamefont {Y.}~\bibnamefont {Xu}}, \bibinfo {author}
  {\bibfnamefont {J.}~\bibnamefont {Zhang}}, \ and\ \bibinfo {author}
  {\bibfnamefont {Y.}~\bibnamefont {Wang}},\ }\href {\doibase
  10.1038/s41563-019-0573-3} {\bibfield  {journal} {\bibinfo  {journal} {Nature
  Materials}\ } (\bibinfo {year} {2020}),\
  10.1038/s41563-019-0573-3}\BibitemShut {NoStop}%
\end{thebibliography}%

\end{document}